\documentclass[prd,preprint,amsmath,amssymb,nofootinbib,eqsecnum]{revtex4-1}
\usepackage{latexsym,ifthen,graphics,color,epsfig,hyperref,mathrsfs}

\newcommand{\ie}{{i.e.}}

\newcommand{\wrt}{with respect to}
\newcommand{\lhs}{left-hand side}

\newcommand{\rhs}{right-hand side}

\newcommand{\naive}{na\"{\i}ve}

\newcommand{\be}{\begin{equation}}
\newcommand{\ee}{\end{equation}}
\newcommand{\bea}{\begin{eqnarray}}
\newcommand{\eea}{\end{eqnarray}}
\newcommand{\beas}{\begin{eqnarray*}}
\newcommand{\eeas}{\end{eqnarray*}}

\newcommand{\bear}{\begin{array}{l}}
\newcommand{\eear}{\end{array}}

\newcommand{\bcf}{\begin{center}\begin{figure}}
\newcommand{\ecf}{\end{figure}\end{center}}

\newcommand{\bct}{\begin{center}\begin{table}}
\newcommand{\ect}{\end{table}\end{center}}

\newcommand{\ds}{\displaystyle}

\newcommand{\eq}[1]{(\ref{eq:#1})}

\newcommand{\eqs}[2]{(\ref{eq:#1}) and~(\ref{eq:#2})}

\newcommand{\sect}[1]{section~\ref{sec:#1}}

\newcommand{\D}{d}

\newcommand{\MomInt}[2]{\int \!\! \frac{d^{#1} #2}{(2\pi)^{#1}} \, }

\newcommand{\measure}[1]{\mathcal{D} #1 \, }
\newcommand{\Fint}[1]{\int \mathcal{D} #1 \,}

\newcommand{\der}[2]{\frac{d #1}{d #2}}

\newcommand{\fder}[2]{\frac{\delta #1}{\delta #2}}
\newcommand{\dfder}[3]{\frac{\delta^2 #1}{\delta #2 \delta #3}}

\newcommand{\Or}{\mathrm{O}}
\newcommand{\order}[1]{\Or \bigl( #1 \bigr)}

\newcommand{\hf}{\frac{1}{2}}

\newcommand{\Tr}{\mathrm{Tr}\,}

\newcommand{\deltahat}[1]{
	\bar{\delta}(#1)
}

\newcommand{\pf}{\mathcal{Z}}
\newcommand{\field}{\phi}
\newcommand{\dfield}{\varphi}
\newcommand{\jbar}{\bar{j}}

\newcommand{\cutoff}{K}

\newcommand{\ep}{C}
\newcommand{\epIR}{D}
\newcommand{\dd}{\dot{\ep}}
\newcommand{\knl}[1]{\cdot {#1}\cdot}

\newcommand{\flow}{\Lambda \partial_\Lambda}
\newcommand{\flowk}{k \partial_k}
\newcommand{\op}{\hat{A}}
\newcommand{\Count}{\Delta}

\newcommand{\Stot}{S^{\mathrm{tot}}}
\newcommand{\Sint}{S}
\newcommand{\SintJ}{T}
\newcommand{\Gtot}{\Gamma^{\mathrm{tot}}}
\newcommand{\clf}{\field^{\mathrm{c}}}

\newcommand{\dual}{\mathcal{D}}
\newcommand{\dualJ}{\mathcal{E}}
\newcommand{\homog}{\mathcal{H}}
\newcommand{\homogpr}{\mathcal{G}}
\newcommand{\homogv}[1]{\mathcal{H}_{#1}}

\newcommand{\classical}[3]{\fder{#1}{\field} \knl{#2} \fder{#3}{\field}} 
\newcommand{\quantum}[2]{\fder{}{\field} \knl{#1} \fder{#2}{\field}} 

\newcommand{\classicalvar}[3]{\fder{#1}{\dfield} \knl{#2} \fder{#3}{\dfield}} 
\newcommand{\quantumvar}[2]{\fder{}{\dfield} \knl{#1} \fder{#2}{\dfield}} 

\newcommand{\classicalJ}[3]{\fder{#1}{J} \knl{#2} \fder{#3}{J}}
\newcommand{\quantumJ}[2]{\fder{}{J} \knl{#1} \fder{#2}{J}}

\newcommand{\classicaljpr}[3]{\fder{#1}{j_g} \knl{#2} \fder{#3}{j_g}}
\newcommand{\quantumjpr}[2]{\fder{}{j_g} \knl{#1} \fder{#2}{j_g}} 

\newcommand{\classicalscrJ}[3]{\fder{#1}{\!\mathscr{J}} \knl{#2} \fder{#3}{\!\mathscr{J}}}
\newcommand{\quantumscrJ}[2]{\fder{}{\!\mathscr{J}} \knl{#1} \fder{#2}{\!\mathscr{J}}} 

\newcommand{\eop}{\mathcal{O}}

\newcommand{\marginalR}{\eop_\mathrm{mar}^{\mathrm{red}}}

\newcommand{\const}{c}

\newcommand{\hepth}[1]{hep-th/#1}
\newcommand{\hepph}[1]{hep-ph/#1}

\newcommand{\condmat}[1]{cond-mat/#1}

\begin{document}

\title{Equivalent Fixed-Points in the Effective Average Action Formalism}

\author{Oliver~J.~Rosten}

\affiliation{Department of Physics and Astronomy, University of Sussex, Brighton, BN1 9QH, U.K.}
\email{O.J.Rosten@Sussex.ac.uk}

\begin{abstract}
	Starting from a modified version of Polchinski's equation, Morris' fixed-point equation for the effective average action is derived. Since an expression for the line of equivalent fixed-points associated with every critical fixed-point is known in the former case, this link allows us to find, for the first time, the analogous expression in the latter case.
\end{abstract}

\maketitle

\section{Introduction}

Exact Renormalization Group (ERG) equations comes in many different guises. The ideology behind Wilson's groundbreaking understanding of renormalization~\cite{Wilson} is most obvious in formulations which explicitly involve some sort of coarse-graining procedure. Roughly speaking, this process---inspired by Kadanoff~\cite{Kadanoff}---involves partitioning a system up into small patches and then averaging over the degrees of freedom within each patch in an appropriate way. A key requirement is that this operation leaves the partition function invariant. As recognized by Wegner, in particular, this allows for ERGs to be formulated in a very general way~\cite{WegnerInv}.

Denoting the approximate inverse size of a patch by $\Lambda$, `the effective scale', we introduce the Wilsonian effective action, $\Stot_\Lambda$. (Where the `tot' is for `total'; we reserve the symbol $S_\Lambda$ for something slightly different.) If the coarse-graining is initiated at the bare scale, $\Lambda_0$, then $\Stot_\Lambda$ incorporates the effects of all fluctuations (be they quantum or statistical) between the bare and effective scales. Working with theories of a single scalar field, $\field$, invariance of the partition function function can be achieved by taking
\be
-\flow e^{-\Stot_\Lambda[\field]} =  \int_p \fder{}{\field(p)} 
	\left\{
	\Psi(p) e^{-\Stot_\Lambda[\field]}
	\right\},
\label{eq:blocked}
\ee
where $\int_p \equiv \MomInt{\D}{p}$ and we understand that $\Psi(p)$, which must depend on $\Stot_\Lambda$~\cite{Wegner_CS,Fundamentals}, encodes the details of the precise blocking procedure of choice (for further details see~\cite{mgierg1,Fundamentals}). Working as we do in momentum space, an infinitesimal reduction of the effective scale amounts to integrating over an infinitesimal shell of momentum modes in the partition function. Let us note that $\Psi$ can be interpreted as implementing an infinitesimal field redefinition~\cite{Wegner_CS,TRM+JL}.

For the purposes of this paper, we will concern ourselves with a choice of $\Psi$ which gives rise to Polchinski's ERG equation~\cite{Pol} or a particular modification thereof~\cite{Ball}. A central ingredient is an ultraviolet (UV) cutoff function, $\cutoff(p^2/\Lambda^2)$, which, for $p^2 \sim \Lambda^2$, should generally be taken to die off faster than any power~\cite{Fundamentals}. In the infrared (IR) $\cutoff(p^2/\Lambda^2)$ is \emph{quasi-local}, meaning that it exhibits an all-orders Taylor expansion, a requirement necessary to ensure that the coarse-graining is performed over suitably local patches~\cite{aprop}. The normalization is chosen such that
\be
	\cutoff(0) = 1.
\label{eq:cutoff-condition}
\ee
It is convenient to split off a piece of the total action which is naturally identified as a regularized kinetic term:
\be
	\Stot_\Lambda[\field] = \hf \field \cdot \ep^{-1}_\Lambda \cdot \field + \Sint_\Lambda[\field],
\label{eq:split}
\ee
where $\field \cdot \ep^{-1}_\Lambda \cdot \field = \int_p \phi(p) \ep^{-1}_\Lambda(p^2) \phi(-p)$
and
\be
	\ep_\Lambda(p^2) = \frac{\cutoff(p^2/\Lambda^2)}{p^2}.
\label{eq:ep}
\ee
Note, though, that in general $S_\Lambda$ can contain additional two-point pieces so it should not be presumed from the form of~\eq{ep} that the theory is necessarily massless. (Indeed, the suggested interpretation of the two-point piece above, whilst usually helpful, can be misleading; for example, we might find a solution to the flow equation such that $S_\Lambda$ subtracts off the $\order{p^2}$ part belonging to the integrand of $\field \cdot \ep^{-1}_\Lambda \cdot \field$~\cite{Wegner_CS,Fundamentals}.)

Defining $\dd_\Lambda \equiv - \Lambda d \ep_\Lambda / d\Lambda$, the flow equations of interest follow from choosing
\be
	\Psi(p) = \hf \dd_\Lambda(p^2) 
	\biggl\{
		\fder{\Sint_\Lambda[\field]}{\field(-p)} - \ep^{-1}_\Lambda(p^2) \field(p)
	\biggr\}
	+ \psi(p),
\label{eq:choice}
\ee
which, upon substitution into~\eq{blocked}, yields
\be
	-\flow \Sint_\Lambda[\field] = \hf \classical{\Sint}{\dd}{\Sint} - \hf \quantum{\dd}{\Sint} 
	+ \psi \cdot \ep^{-1}_\Lambda \cdot \field
	+ \psi \cdot \fder{\Sint}{\field} - \fder{}{\field} \cdot \psi,
\label{eq:Pol-psi}
\ee
where $\psi \cdot \delta/\delta \field = \int_p \psi(p)\, \delta/\delta \field(p)$ and we have dropped the dependencies of $S$ on the \rhs\ for brevity.
Given our choice of $\Psi$, \eq{choice}, $\psi$ encodes the residual freedom to perform 
 an additional field redefinition along the flow. In this paper, we will make one of two choices: either $\psi(p)=0$, recovering the Polchinski equation, or $\psi(p) = -\eta \, \field(p)/2$, yielding the modified Polchinski equation of~\cite{Ball}. In the latter case, 
we can choose $\eta$ such that the corresponding field redefinition ensures canonical normalization of the kinetic term. Denoting the field strength renormalization by $Z$, we therefore identify
\be
	\eta = \Lambda \der{\ln Z}{\Lambda}
\ee
as the anomalous dimension of the field. 

Our focus up until now has been on flow equations which describe how the Wilsonian effective action changes as the effective scale---which plays the role of a UV cutoff---is lowered. However, 
there is a different approach that can be taken based instead on a flow \wrt\ an IR cutoff, which we will denote by $k$. In this case the object of interest is the effective average action, $\Gamma_k$: the IR regularized generator of the one-particle irreducible (1PI) pieces of the Green's functions. There are several different derivations of the flow equation for $\Gamma_k$ on the market (for reviews focusing on this formalism see~\cite{Wetterich-Rev,Delamotte-Rev,JMP-Review,Gies-Rev}). 

Wetterich~\cite{Wetterich-1PI} considered adding an IR cutoff function to the bare action, such that the partition function in the presence of a source becomes $k$-dependent:
\be
	\pf_k[J] = \Fint{\field} e^{-\Stot_{\Lambda_0}[\field] -\hf \field \cdot  R_k \cdot \field + J \cdot \field}.
\ee
In order to implement an IR regularization, the function $R_k(p^2)$ satisfies $\lim_{p^2/k^2 \rightarrow 0} R_k(p^2) > 0$. Moreover, $\lim_{k^2/p^2 \rightarrow 0} R_k(p^2) = 0$ so that the regularization disappears as the IR scale is sent to zero (Wetterich also gives a third condition on the regulator~\cite{Wetterich-1PI}). 
The regulator term has a natural interpretation as a $k$-dependent mass term and, as such, the flow equation obtained by differentiating \wrt\ $k$ (and performing the appropriate Legendre transform) is often considered to belong to the family of Callan-Symanzik style flows. 

However, there is an alternative way of deriving the flow equation for $\Gamma_k$. As recognized by Morris~\cite{TRM-ApproxSolns}, if we identify $k$ with $\Lambda$, then $\Gamma_\Lambda$ is related by a Legendre transform to $S_\Lambda$, \emph{so long as the latter satisfies the Polchinski equation}. At first sight it might seem rather strange that $\Lambda$ can play the role of both a UV and an IR cutoff. But, recalling that degrees of freedom between $\Lambda_0$ and $\Lambda$ have been integrated out, this is perfectly natural: $\Lambda$ is a UV cutoff for the unintegrated modes but an IR cutoff for the integrated ones.

Let us emphasise that by linking $\Gamma_\Lambda$ to $S_\Lambda$ in this way, the former inherits the power of the Wilsonian approach. However,
this relationship between the effective average action and the Wilsonian effective action begs an obvious question: what if the latter obeys a flow equation other than the Polchinski equation?
If, for this new flow equation, we take the same boundary condition \ie\ the same bare action, then clearly Wetterich's approach---and hence the flow equation for $\Gamma_k$---is unchanged. However, the bare action is not always something we are free to choose. In particular, if we are interested in scale-invariant theories corresponding to critical fixed-points, then the action is something for which we should \emph{solve}. 

The recipe for doing this is as follows. First, we must work with the modified Polchinski equation, with $\psi =  -\eta \, \field(p)/2$. This will allow us to conveniently find critical theories with a non-vanishing anomalous dimension. Next, we scale the canonical dimension out of the field and coordinates using the effective scale, $\Lambda$, allowing us to formulate the fixed-point condition for the Wilsonian effective action simply as
\be
	\flow \Sint_\star[\dfield] = 0,
\ee
where $\dfield$ is the field after rescaling to dimensionless variables and we use a $\star$ to denote fixed-point quantities.  Our aim now is to \emph{define} a new $\Gamma_k$, which is a functional of a new field $\Phi$, such that  if we scale out the canonical dimensions using the IR scale, $k$, then the above fixed-point condition translates to
\be
	\flowk \Gamma_\star[\Phi] = 0.
\label{eq:Gamma-fp}
\ee

It might seem strange that a $\Gamma_k$ needs to be specially cooked up to satisfy this condition. The reason can be understood as follows. We start with a fixed-point, $\Sint_\star[\dfield]$. This is the most primitive object in our construction. Any quantity we construct from $\Sint_\star[\dfield]$ is, of course, automatically derived from a fixed-point. However, one can easily imagine constructing any number of objects for which this is far from obvious (without prior knowledge). Our task, then, is to construct a $\Gamma_k$ such that, simply by inspection, it is obvious whether or not it derives from a fixed-point, $\Sint_\star[\dfield]$. We do this by arranging things such that, if $\Gamma_k$ \emph{is} derived from a fixed-point, then there are variables for which~\eq{Gamma-fp} is satisfied.

Actually, the equation satisfied by $\Gamma_\star[\Phi]$ in this scenario was deduced long ago by Morris~\cite{TRM-Deriv}, using general considerations. (Specifically, see equation~(5) of~\cite{TRM-Deriv} specialized to fixed-points; note that in the current paper we will not bother to factor out the $\D$-dimensional solid angle from our analogous expressions.) However, in this paper we will derive the equation from first principles. This serves two purposes: one the one hand, it will clarify the relationship between this flow equation and the modified Polchinski equation; on the other, it will allow us to immediately deduce a new result.
  
This new result pertains to the line of equivalent fixed-points associated with each critical fixed-point, where equivalent fixed-points are those related to each other by quasi-local field redefinitions.  Essentially, the physics encoded by a fixed-point is unchanged by changing the normalization of the field, and this invariance manifests itself as a dependence of each critical fixed-point on an unphysical parameter, to be denoted by $b$. In particular, given a critical fixed-point, $\Sint_\star$, and some reference value of $b$, say $(b_0)$, then it was shown in~\cite{Fundamentals,OJR-Pohl} that, for real parameter $a$,
\be
	e^{a\hat{\Count}}  \Sint_\star[\dfield](b_0)
	=
	\Sint_\star[\dfield](b_0+a),
	\quad
	\mathrm{with}
	\quad
	b_0 + a = b
\label{eq:line}
\ee
where it assumed that no singularities are encountered between $b_0$ and $b$ and
\be
	\hat{\Count} \equiv 
	\hf \dfield \cdot \fder{}{\dfield} 
	+ \cutoff \cdot \fder{}{\cutoff}.
\ee  
 (Note that we will indicate operators by a hat.) Each such line of fixed-points is generated by a marginal, redundant operator given by
\be
	\marginalR[\dfield](b_0) = \der{\Sint_\star[\dfield](b_0+a)}{a}\biggr\vert_{a=0}.
\ee

Now, given that in this paper a link is established between $\Sint_\star$---understood as a solution of the modified Polchinski equation---and $\Gamma_\star$, we can use~\eq{line} to derive an expression for the line of equivalent fixed-points in the effective average action formalism.

The results of this paper thus pertain to structural aspects of the ERG. This area of study is rather underdeveloped compared to applications~\cite{Wetterich-Rev,Delamotte-Rev,Fundamentals} of the formalism. However, it is reasonable to hope that an increased understanding of the workings of the ERG will lead to developments in its practical use. Indeed, the recent discovery of the explicit expression for the line of fixed-point given in~\eq{line} led directly to an extension of Pohlmeyer's theorem~\cite{OJR-Pohl}. It is worth mentioning in this context that whilst structural considerations---dating back to~\cite{Wegner_CS}---have generally utilized the Wilsonian effective action, the bulk of modern applications use the effective average action. 
With an eye on the future effectiveness of the formalism, it thus make sense to translate developments in one approach into the other---as is done in what follows.

The rest of this paper is arranged as follows.  In \sect{FEs} we show how to derive a flow equation for the effective average action in two different scenarios. First of all, we will re-derive the standard flow equation for $\Gamma_k$ by starting from the Polchinski equation. This analysis will be seen to be reminiscent of Ellwanger's~\cite{Ellwanger-1PI}. Armed with the lessons learnt from this, we will adapt what we have done to the case of the modified Polchinski equation in \sect{modified}. In fact, we will not give a general treatment but rather will work only at fixed-points, re-deriving Morris' equation of~\cite{TRM-Deriv}. This result will be sufficient to find an expression for the line of equivalent fixed in the effective average action formalism, which will be done in \sect{line}. The analysis of this paper is, in places, rather involved. Consequently, the first part of the conclusion is devoted to giving an overview of the main steps. The conclusion closes with some remarks on generalizations and possible future work.

\section{Flow Equations}
\label{sec:FEs}

Throughout this paper (in which we work in $\D$-dimensional Euclidean space), it will be useful to consider allowing the action to depend not just on $\field$, but also on an external field, $J$. In this case, a perfectly good flow equation follows simply by replacing
$S_\Lambda[\field]$ in~\eq{Pol-psi} with $T_\Lambda[\field,J]$, where $T_\Lambda[\field,0]=S_\Lambda[\field]$. If we choose the boundary condition to the flow to be
\be
	\lim_{\Lambda \rightarrow \Lambda_0} 
	\Bigl(
		T_\Lambda[\field,J] - S_\Lambda[\field]
	\Bigr)
	= -J \cdot \field,
\label{eq:bc}
\ee
then the $J$-dependence of $T[\field,J]$ is such that the standard correlation functions (\ie\ those obtained from derivatives of $W[J]$) can be picked out (in a manner to be made precise below).

\subsection{The Polchinski Equation}
\label{sec:Pol}

In this section we will focus on the case $\psi(p) = 0$. As noted in~\cite{TRM-ApproxSolns,Salmhofer} the Polchinski equation can be linearized. Recalling that $\Lambda$ and $k$ are our UV and IR scales, respectively, we start by constructing the following object
\be
	\cutoff^\Lambda_k(p^2) \equiv \cutoff(p^2/\Lambda^2) - \cutoff(p^2/k^2),
\label{eq:Cutoff-UV-IR}
\ee
which we note effectively has support only in the range $k^2 \lesssim p^2 \lesssim \Lambda^2$. In turn, this leads us to define
\be
	\ep^\Lambda_k(p^2) \equiv \frac{\cutoff^\Lambda_k(p^2)}{p^2} = \ep_\Lambda(p^2) - \ep_k(p^2)
\label{eq:ep-UV-IR}
\ee
and now to introduce the operator
\be
	\op_k^\Lambda \equiv
	\hf
	\classical{}{\ep^\Lambda_k}{}.
\label{eq:op-reg}
\ee

In the current scenario---where the Wilsonian effective action satisfies the Polchinski equation---there is a simple relationship between $\Sint_k$ and $\Sint_\Lambda$ and also $\SintJ_k$ and $\SintJ_\Lambda$:
\be
	-\Sint_k[\field] = \ln \Bigl( e^{\op^\Lambda_k} e^{-\Sint_\Lambda[\field]} \Bigr), 
	\qquad
	-\SintJ_k[\field,J] = \ln \Bigl(e^{\op^\Lambda_k} e^{-\SintJ_\Lambda[\field,J]} \Bigr).
\label{eq:Sint_k}
\ee
This can be checked by first noticing that the Polchinski equation implies
\be
	\flow \Sint_k[\field] = 0,
	\qquad
	\flow \SintJ_k[\field,J] = 0,
\label{eq:Sint_k-flow}
\ee
and then taking the limit $\Lambda \rightarrow k$ in~\eq{Sint_k}. It is thus apparent that the Polchinski equation, which is non-linear in $\Sint_{\Lambda}$, implies a linear equation in $\Sint_k$.

Consider now the limit $k \rightarrow 0$ in~\eq{Sint_k}. From~\eq{ep-UV-IR} it is apparent that $\cutoff^\Lambda_0(p^2) =  \ep_\Lambda(p^2)$. However, taking this limit in~\eq{Sint_k} is subtle due to the possible appearance of IR divergences.
Nevertheless, if we assume that the limit $k \rightarrow 0$ is just the \naive\ one, then $\SintJ_{k=0}$ generates the connected correlation functions according to~\cite{Fundamentals}:
\be
	G(p_1,\ldots,p_n)
	\deltahat{p_1+\cdots+p_n}
	=
	-
	\left.\fder{}{J(p_1)} \cdots \fder{}{J(p_n)} \SintJ_{k=0}[0,J]\right\vert_{J=0},
\ee
where $\deltahat{p} \equiv (2\pi)^{\D} \delta^{\D}(p)$.
Consequently, we interpret
\be
	\SintJ_{k}[0,J] = -W_k[J]
\label{eq:W_k}
\ee	
as the generator of IR cutoff correlation functions.

Since $\SintJ_k[\field,J]$ is independent of $\Lambda$, we can evaluate it at any $\Lambda$ of our choosing and get the same result. With this in mind, let us do so at the bare scale, and use the boundary condition~\eq{bc}. We find that~\cite{Fundamentals,HO-Remarks}:
\be
	\SintJ_k[\field,J] 
	=
	-\ln 
	\Bigl(
		e^{\op^{\Lambda_0}_k} e^{-\Sint_{\Lambda_0}[\field] + J \cdot \field}
	\Bigr)
	=
	e^{J \cdot \ep^{\Lambda_0}_k \cdot \delta /\delta\field}
	\Sint_k[\field]
	-J\cdot \field - \hf J \cdot \ep^{\Lambda_0}_k \cdot J,
\label{eq:dualJ-dual}
\ee
from which it follows that
\be
	W_k[J] =  \hf J \cdot \ep^{\Lambda_0}_k \cdot J - \Sint_k[\ep^{\Lambda_0}_k J].
\label{eq:W-dual}
\ee
This result enables us to obtain a flow equation for the effective average action \ie\ the generator of IR cutoff 1PI diagrams. Anticipating that we will allow $J$ to depend on $k$, we start by 
noticing from~\eq{Sint_k} that
\be
	\biggl(
		\left.\flowk\right\vert_J  
		+ J \cdot \dd^{\Lambda_0}_k D^{\Lambda_0}_k \cdot \fder{}{J}
	\biggr)
	\Sint_k[\ep^{\Lambda_0}_k J] = 
	-\hf\classicalJ{\Sint_k}{\dot{D}^{\Lambda_0}_k}{\Sint_k} 
	+ \hf\quantumJ{\dot{D}^{\Lambda_0}_k}{\Sint_k},
\label{eq:dual_k_CJ-flow}
\ee
where we have defined
\be
	D^{\Lambda_0}_k(p^2) \equiv  \bigl[\ep^{\Lambda_0}_k(p^2) \bigr]^{-1}.
\label{eq:D}
\ee
and we understand  $\dd^{\Lambda_0}_k \equiv -k d \ep^{\Lambda_0}_k /d k$ (and similarly for $\dot{D}^{\Lambda_0}_k$).
Substituting~\eq{W-dual} into~\eq{dual_k_CJ-flow} it is simple to check that, up to a discarded
vacuum energy term,
\be
	\flowk W_k[J]
	=
	\hf \classicalJ{W_k}{\dot{D}^{\Lambda_0}_k}{W_k}
	+\hf \quantumJ{\dot{D}^{\Lambda_0}_k}{W_k}.
\label{eq:flow-W}
\ee

To derive the flow equation for the effective average action, we perform the usual Legendre transform, for which we follow Weinberg's treatment~\cite{WeinbergII}.
First of all, we introduce the classical field in the presence of the source (and an IR regulator):
\be
	\clf_J(p) \equiv \fder{W_k[J]}{J(-p)}.
\label{eq:dWdJ}
\ee
Next we adjust $J(p)$ to a specific $J_\field(p)$ such that the classical field takes on a prescribed form $\clf_J(p) = \clf(p)$. Then we define
\be
	\Gtot_k[\clf] \equiv J_\field \cdot \clf -W_k[J_\field].
\label{eq:Gamma_k}
\ee
Differentiating $\Gtot_k$ \wrt\ $\clf$ and using~\eq{dWdJ} yields
\be
	\fder{\Gtot_k[\clf]}{\clf(-p)} = J_\field(p).
\label{eq:dGdvarphi}
\ee
From~\eqs{dWdJ}{dGdvarphi} it follows, in the standard way~\cite{Wetterich-1PI,WeinbergII}, that
\be
	\int_q \dfder{\Gtot_k[\clf]}{\clf(p)}{\clf(q)} \dfder{W_k[J_\field]}{J_\field(-q)}{J_\field(-p')}
	= \deltahat{p-p'}.
\label{eq:usual}
\ee
Plugging~\eq{Gamma_k} into the \lhs\ of~\eq{flow-W} and using~\eqs{dWdJ}{usual} on the \rhs\
yields
\be
	\flowk\vert_{J_\field}
	\bigl(
		J_\field \cdot \clf -\Gtot_k[\clf]
	\bigr)
	=
	\hf \clf \cdot \dot{D}^{\Lambda_0}_k \cdot \clf 
	+ \hf \Tr 
	\biggl[
	\dot{D}^{\Lambda_0}_k
	\biggl(
		\dfder{\Gtot_k}{\clf}{\clf}
	\biggr)^{-1}
	\biggr].
\ee
Substituting for $J$ on the \lhs\ using~\eq{dGdvarphi}, it is apparent that we can drop the resulting term if we take derivative \wrt\ $k$ to be performed at constant $\clf$. If we additionally define
\be
	\Gamma_k[\clf] \equiv
	\Gtot_k[\clf] -
	\hf \clf \cdot D^{\Lambda_0}_k \cdot \clf,
\label{eq:Gamma}
\ee
then, dropping another vacuum term, we arrive at the standard equation~\cite{TRM-ApproxSolns,Wetterich-1PI,Ellwanger-1PI}
\be
	-\flowk \Gamma_k[\clf]
	=
	\hf \Tr
	\Bigl\{
	 \dot{D}^{\Lambda_0}_k
	 \Bigl[
	 	D^{\Lambda_0}_k + \Gamma_k^{(2)}
	 \Bigr]^{-1}
	 \Bigr\},
\label{eq:Morris-1PI}
\ee
where $\Gamma^{(2)}_k \equiv \delta^2 \Gamma_k /\delta \clf \, \delta \clf$.

Before moving on, let us re-express $\Gamma_k$ in terms of $S_k$. This can be achieved by substituting~\eq{W-dual} into~\eq{Gamma_k} and finally using~\eq{Gamma}. Setting $\chi \equiv \ep^{\Lambda_0}_k J_\phi$, 
the result is that
\be
	\Gamma_k[\clf] = S_k[\chi]
	-
	\hf
	\bigl(
		\clf  -\chi
	\bigr)
	\cdot
		D^{\Lambda_0}_k
	\cdot
	\bigl(
		\clf  - \chi
	\bigr),
\ee
recovering a result due to Morris~\cite{TRM-ApproxSolns}.

For most applications, the bare scale $\Lambda_0$ is now sent to infinity. This does not actually amount to an assumption of renormalizability, as we will discuss in a moment. First, though, 
let us note that $\cutoff^\infty_k(p^2)$ effectively has support for $k^2 \lesssim p^2 < \infty$ and so can be interpreted as an IR cutoff function. Now, as in the work of Morris~\cite{TRM-ApproxSolns}, this cutoff function appears multiplicatively, in the sense that we understand its appearance as a multiplicative modification of the canonical kinetic term: $ p^2 \rightarrow [\cutoff^\infty_k(p^2)]^{-1} p^2 = \epIR^\infty_k(p^2)$. This is to be contrasted with Wetterich's approach where, as we have seen, the IR cutoff appears in an additive fashion: $p^2 \rightarrow p^2 + R_k(p^2)$. Were we to redefine $\Gamma_k[\clf] \rightarrow \Gamma_k[\clf] + \hf \int_p \clf(p) \clf(-p) p^2$, then the equation of~\cite{Wetterich-1PI} follows from replacing $D^\infty_k$ with $R_k$ in~\eq{Morris-1PI}. Either way, the fact that both terms on the \rhs\ of~\eq{Morris-1PI} appear multiplied by $\dot{D}^{\Lambda_0}_k$ is important: this differentiated object effectively has support only for $p^2 \sim k^2$ and so serves as both an IR and a UV regulator, in this context. Therefore, even if we send $\Lambda_0 \rightarrow \infty$, the flow equation~\eq{Morris-1PI} is regularized. Solutions of this equation follow from specifying a boundary condition at some reference scale $k = k_0$ and integrating along the flow. Renormalizable theories can be picked out as those solutions for which (in variables rendered dimensionless using $k$), there is no explicit dependence on $k_0$.

\subsection{The Modified Polchinski Equation}
\label{sec:modified}

In this section we will treat the modified version of the Polchinski mentioned around~\eq{choice}. In \sect{MPE-FE}, we will give the explicit form of the flow equation. It will be noticed that if we attempt to introduce an IR cutoff function in a similar manner to~\eq{Sint_k}, then the resulting objects do not satisfy linear equations 
as they did previously. Instead, we will recall the objects derived from $\Sint$ and $\SintJ$ which \emph{do}  satisfy linear equations~\cite{Trivial,Fundamentals} and give a recipe for constructing a flow equation for the effective average action.

However, rather than dealing with a full flow equation for $\Gamma_k$, we will instead focus on fixed-points, about which some useful facts are recalled in \sect{MPE-FPs}. Armed with the lessons learnt, in \sect{MPE-First} we attempt to construct a $\Gamma_k$. However, part way through the process, it becomes apparent that we have no hope of satisfying the convenient condition~\eq{Gamma-fp} and so we abort. But at this stage it is clear how we \emph{can} introduce a $\Gamma_k$ which has the desired property, and this is done in \sect{MPE-Second}.

\subsubsection{The Flow Equation and its Linearization}
\label{sec:MPE-FE}

In this section we return to~\eq{Pol-psi} and, instead of taking $\psi(p)=0$, take $\psi(p) = -\eta \, \field(p)/2$. Moreover, (to start with) we will work in variables which have been rendered dimensionless by using the effective scale, $\Lambda$. First of all, we define $\tilde{p} \equiv p/\Lambda$. Now, given some field $X(p)$, 
with (canonical) dimension $[X(p)]$, we introduce the dimensionless field $x(\tilde{p}) = X(p) \Lambda^{-[X(p)]}$. Therefore, we take $\dfield(\tilde{p}) = \field(p) \Lambda^{(\D+2)/2}$ and
$j(\tilde{p}) = J(p) \Lambda^{(d-2)/2}$. Notice that the functional derivative $\delta /\delta X(p)$ has dimension $[X(p)]- \D$, consistent with $\delta X(p) /\delta X(q) = \deltahat{p-q}$. Since we want everything in our flow equation to be dimensionless, we take
\be
	\fder{}{x(\tilde{p})} = \Lambda^{\D-[X(p)]} \fder{}{X(p)}.
\label{eq:fder-dimless}
\ee
Henceforth, we will drop the tildes: whether or not dimensionless momenta are being used can be deduced from the context. Finally, we introduce an arbitrary scale, $\mu$ and use it to define the `RG-time' $t = \ln \mu/\Lambda$. In dimensionless variables, the flow equation~\eq{Pol-psi} (extended to allow for source-dependence of the action) reads:
\be
	\bigl(\partial_t -\hat{D}^- -\hat{D}^j \bigr) \SintJ_t[\dfield,j]
	= \classicalvar{\SintJ}{\cutoff'}{\SintJ}
	-\quantumvar{\cutoff'}{\SintJ}
	-\frac{\eta}{2} \dfield \cdot \ep^{-1} \cdot \dfield,
\label{eq:Ball-source}
\ee
where $\cutoff'(p^2) \equiv d \cutoff(p^2)/d p^2$, we understand $\partial_t$ to act under the integrals (\ie\ we do not differentiate the dimensionless momenta; for a further discussion see~\cite{Fundamentals}) and take
\be
	\hat{D}^{\pm} = \int_p
	\biggl[
		\biggl( \frac{\D+2 \pm \eta}{2} + p \cdot \partial_p \biggr) \dfield(p)
	\biggr]
	\fder{}{\dfield(p)},
	\qquad
	\hat{D}^j = 
	 \int_p
	\biggl[
		\biggl( \frac{\D- 2 + \eta}{2} + p \cdot \partial_p \biggr) j(p)
	\biggr]
	\fder{}{j(p)}.
\ee
Of course, the source-independent version follows simply from replacing $\SintJ_t[\dfield,j]$ with $\Sint_t[\dfield]$, after which the $\hat{D}^j$ term can be dropped.

Attempting to mimic the analysis of the previous section, it would seem natural to define, along the lines of~\eq{Sint_k}, two objects
\be
	-\dual_{t,\kappa}[\dfield] \equiv \ln \Bigl( e^{\op^1_{\kappa}} e^{-\Sint_t[\dfield]} \Bigr), 
	\qquad
	-\dualJ_{t,\kappa}[\dfield,j] \equiv \ln \Bigl(e^{\op^1_{\kappa}} e^{-\SintJ_t[\dfield,j]} \Bigr),
\label{eq:dual-rescale}
\ee
where $\kappa \equiv k/\Lambda$ and
\be
	\op^1_{\kappa}
	=
	\hf
	\int_p
	\fder{}{\dfield(p)}
	\frac{\cutoff(p^2) - \cutoff(p^2/\kappa^2)}{p^2}
	\fder{}{\dfield(-p)}.
\ee 
Annoyingly, the presence of the final term on the \rhs\ of~\eq{Ball-source} complicates the analysis of the previous section. Not only do $\dual_k$ and $\dualJ_k$ no longer reduce, respectively, to $\Sint_k$ and $\SintJ_k$ but,
as pointed out in~\cite{OJR-Pohl,Fundamentals}, the flow equation~\eq{Ball-source} does not even imply
a linear equation for $\dual_k$ and $\dualJ_k$.

However, the flow equation does linearize if we make the tacit assumption that the objects defined without ever introducing IR regularization,
\be
	-\dual_t[\dfield] \equiv -\ln \Bigl( e^{\op} e^{-\Sint_t[\dfield]} \Bigr), 
	\qquad
	-\dualJ_t[\dfield,j] \equiv -\ln \Bigl(e^{\op} e^{-\SintJ_t[\dfield,j]} \Bigr),
\label{eq:dual-dimless}
\ee
exist and are sufficiently well behaved.\footnote{For $\dual$, at any rate, this is very reasonable. For theories sitting at a critical fixed-point which, being massless, potentially have IR problems, the vertices of $\dual_\star$ are better behaved than those of the correlation functions (\ie\ $\dualJ_\star$) by a power of momentum squared on each leg.} The meaning of the second condition will become clear below.
Note that we take 
\be
	\op = \int_p\fder{}{\dfield(p)} \frac{\cutoff(p^2)}{p^2} \fder{}{\dfield(-p)}, 
\ee
where we recall that $p$ has been rendered dimensionless using $\Lambda$. Computing the flow of $\dual_t$ and $\dualJ_t$ we find that~\cite{Trivial,Fundamentals}
\be
	\left(
		\partial_t - \hat{D}^+ -\frac{\eta}{2} \dfield \cdot \ep^{-1} \cdot \dfield
	\right) 
	e^{-\dual_t[\dfield]}
	= 0,
	\qquad
	\left(
		\partial_t - \hat{D}^+ - \hat{D}^j -\frac{\eta}{2} \dfield \cdot \ep^{-1} \cdot \dfield
	\right) 
	e^{-\dualJ_t[\dfield,j]}
	= 0.
\label{eq:dualJ-DimlessFlow}
\ee

The game now is as follows. 
\begin{enumerate}
	\item Look for solutions to the two equations of~\eq{dualJ-DimlessFlow}.
	In the source-dependent case, the solution of interest 
	must be consistent with the boundary condition~\eq{bc}. 
	Once we have found these solutions, we can then relate $\dualJ_t[\dfield,j]$ to $\dual_t[\dfield]$. 
	
	\item Define appropriate IR regularized versions of these objects, which we will denote 
	by $\dualJ_{t,\kappa}[\dfield,j]$ and $\dual_{t,\kappa}[\dfield]$. Noting that $\dualJ_t[0,j]$ has 
	been 
	shown in the past to generate the connected correlation functions~\cite{Fundamentals},
	we therefore identify $\dualJ_t[0,j] = -W_{t,\kappa}[j]$, which is the analogue of~\eq{W_k}.
		
	\item Use the relationship between $\dualJ_t[\dfield,j]$ and $\dual_t[\dfield]$ to find the relationship 
	between $\dualJ_{t,\kappa}[\dfield,j]$ and $\dual_{t,\kappa}[\dfield]$, which will lead to an 
	equation analogous to~\eq{dualJ-dual}. 
	
	\item Perform the steps leading to~\eq{flow-W} and ultimately to derive the flow equation for 
	$\Gamma_k$ appropriate to the modified Polchinski equation.
\end{enumerate}
However, rather than doing this in full, we instead restrict our interest to critical fixed-points, leaving a general analysis for the future. 

\subsubsection{Critical Fixed-Points}
\label{sec:MPE-FPs}

By focusing on critical fixed-points (for which we recall that $\eta_\star <2$), we can exploit the facts that 
we know both the form of the flow equation for which we are aiming and
the relationship [given~\eq{bc}] between $\dualJ_t[\dfield,j]$ and $\dual_t[\dfield]$~\cite{OJR-Pohl,Fundamentals}:
\be
	\dualJ_\star[\dfield,j] = e^{\jbar \cdot \delta / \delta \dfield} \, \dual_\star[\dfield]
	- \jbar \cdot \rho \cdot \dfield - \hf \jbar \cdot \rho \cdot \jbar,
\label{eq:FP-Relationship}
\ee
where $\jbar(p) \equiv j(p)/p^2$ and
\be
	\rho(p^2) \equiv 
	\ep^{-1}(p^2)
	- p^{2(1+ \eta_\star/2)}  \int_0^{p^2} dq^2
	\left[
		\frac{1}{\cutoff (q^2)}
	\right]'
	q^{-2(\eta_\star/2)}.
\label{eq:rho}
\ee
Given that the cutoff function should be quasi-local, it follows that $\rho(p^2)$ is quasi-local, with the expansion starting at $\order{p^2}$.
For what follows, it will be helpful to define
\be
	\overline{\rho}(p^2) \equiv  \rho(p^2) /p^2 = 1 + \order{p^2}.
\label{eq:sigma-expansion}
\ee

Before moving on, it will be useful to recall  the solution for $\dual_\star[\dfield]$:
\be
	\dual_\star[\dfield] = \homog[\dfield] + \hf \dfield \cdot h \cdot \dfield,
\label{eq:homog}
\ee
where 
\be
	h(p^2) = -\const_{\eta_\star} p^{2(1+\eta_\star/2)}
	+ \rho(p^2),
	\qquad
	\const_{\eta_\star} = 
	\left\{
		\begin{array}{ll}
			1, \ & \eta_\star = 0
		\\
			0, \ & \eta_\star <2, \ \neq 0
		\end{array}
	\right.
\label{eq:h}
\ee
and $\homog$ is a polynomial of the field with vertices that transform homogeneously with momenta. 
(The $c_{\eta_\star}$ are chosen so that $h$ has no contributions that transform in the same way as the vertices of $\homog$.)
To be precise:
\be
	\homog[\dfield] =
	\sum_n \frac{1}{n!} \int_{p_1,\ldots,p_n} \homogv{n}(p_1,\ldots,p_n) \, \dfield(p_1) \cdots \dfield(p_n)
	\deltahat{p_1+\cdots+p_n}
\label{eq:Homog-solution}
\ee
where, for scaling parameter $\xi$,
\be
	\homogv{n}(\xi p_1,\ldots, \xi p_n) = \xi^r \homogv{n}(p_1,\ldots, p_n),
	\qquad
	r = \D - n \frac{\D - 2 - \eta_\star}{2}.
\label{eq:r}
\ee
For what follows, it will be convenient to define
\be
	\homogpr[\dfield] \equiv \homog[\dfield] - 
	\frac{c_{\eta_\star}}{2} \int_p \dfield(p) \dfield(-p) p^{2(1+ \eta_\star/2)},
\label{eq:homogpr}
\ee
from which we have that
\be
	\dual_\star[\dfield] = \homogpr[\dfield] + \hf \dfield \cdot \rho \cdot \dfield.
\label{eq:dual-G}
\ee
Notice from~\eq{h} that $\homog$ and $\homogpr$ only differ when $\eta_\star =0$. Treating the $\eta_\star=0$ case differently from the rest will be seen to be necessary in order to ensure the correct $k \rightarrow 0$ limit of the correlation functions.

\subsubsection{The First Attempt}
\label{sec:MPE-First}

In this section, we will look what happens if we take the obvious choice for $\dual_{t,\kappa}[\dfield] $
and $\dualJ_{t,\kappa}[\dfield,j]$. As will be seen, the results are not desirable, but understanding why this is the case will enable us to refine our approach. With this in mind, let us make the following
indentifications, along the lines of~\eq{dual-rescale}:
\be
	-\dual_{t,\kappa}[\dfield] = \ln \Bigl(e^{\op_{\kappa}} e^{-\dual_t[\dfield] }\Bigr),
	\qquad
	-\dualJ_{t,\kappa}[\dfield,j] = \ln \Bigl( e^{\op_{\kappa}} e^{-\dualJ_t[\dfield,j]}\Bigr),
\label{eq:dual_k}
\ee
where
\be
	\op_{\kappa}
	=
	-
	\hf
	\int_p
	\fder{}{\dfield(p)}
	\frac{\cutoff(p^2/\kappa^2)}{p^2}
	\fder{}{\dfield(-p)}
\ee
and we tacitly assume that operating with $e^{\op_{\kappa}}$ makes sense.
Our earlier assumption that $\dual$ and $\dualJ$ are `sufficiently well behaved' amounts to assuming that the $k \rightarrow 0$ limit of the above equations is the na\"{\i}ve limit \ie\ $\lim_{k \rightarrow 0} \dual_{t,\kappa}[\dfield] = \dual_{t}[\dfield]$, and similarly for $\dualJ[\dfield,j]$.

Let us now specialize to a fixed-point and substitute~\eq{FP-Relationship} into the second equation of~\eq{dual_k} to give:
\be
	\dualJ_{\star,\kappa}[\dfield,j] = 
	e^{\jbar \cdot (1-\rho \, \ep_{\kappa} ) \cdot \delta / \delta \dfield} 
	\, \dual_{\star,\kappa}[\dfield]
	- \jbar \cdot \rho \cdot \dfield - \hf \jbar \cdot \rho \bigl(1-\rho \, \ep_{\kappa}\bigr) \cdot \jbar.
\label{eq:dualJ-dual-fp-relation}
\ee
Notice that
\[
	\jbar \cdot (1-\rho \, \ep_{\kappa} ) \cdot  \fder{}{\dfield}
	=
	\int_p 
	j(p) 
	\biggl[
		\frac{1 - \overline{\rho}(p^2) \cutoff(p^2 /\kappa^2)}{p^2} 
	\biggr]
	\fder{}{\dfield(p)}	
\]
where, crucially, the piece in square brackets is quasi-local (for $\kappa>0$) on account of~\eqs{cutoff-condition}{sigma-expansion}.
Our aim now is to use the relationship~\eq{dualJ-dual-fp-relation}---which we note is reminiscent of~\eq{dualJ-dual}---to derive a flow equation for $\Gamma_k$ which, as before, will be related to $\dualJ_{\star,\kappa}[0,j]$ by a Legendre transform. However, as emphasised before, we would like to set things up in such a way that, when using the appropriate variables, we can write the fixed-point condition for $\Gamma_k$ as $\flowk \Gamma_\star = 0$. So, rather than immediately following the steps which led to~\eq{flow-W}, let us instead consider $\dualJ_{\star,\kappa}[0,j]$ more carefully.

If we substitute~\eq{homog} into the first equation of~\eq{dual_k} then we find that
\begin{align}
\nonumber
	e^{-\dual_{\star,\kappa}[\dfield]} & =
	  e^{\op_{\kappa}}
	  e^{-\homogpr[\dfield] - \hf \dfield \cdot \rho \cdot \dfield }
\\
	& = e^{-\hf \dfield \cdot \rho / (1-\rho \, \ep_{\kappa}) \cdot \dfield }
		\exp
		\biggl(
			-\hf 
			\fder{}{\tilde{\dfield}} 
			\cdot
				\frac{ \ep_{\kappa}}{1 - \rho \, \ep_{\kappa}}
			\cdot
			\fder{}{\tilde{\dfield}}
		\biggr)
		e^{-\homogpr[\tilde{\dfield}]},
\label{eq:dual_star_k-expr}
\end{align}
where $\tilde{\dfield} = \dfield/ (1-\rho \ep_{\kappa})$. This result can be most readily be seen from a diagrammatic perspective. Taking the logarithm on both sides of~\eq{dual_star_k-expr}, $\dual_{\star,\kappa}[\dfield]$ comprises all connected diagrams built out of vertices of $\homogpr$ and the two-point vertex $\rho$~\cite{Fundamentals}. If we commute $\hf \dfield \cdot \rho \cdot \dfield$ to the left on the first line of~\eq{dual_star_k-expr} then the vertex $\rho$ can appear in one of three ways: as a diagram on its own, as a dressing of every external leg or as a dressing of every internal line. Summing up these contributions gives the second line of~\eq{dual_star_k-expr}. We will use this trick---which can, of course, be demonstrated without recourse to diagrammatics---throughout this paper.
Using~\eq{dual_star_k-expr} in~\eq{dualJ-dual-fp-relation} it follows that
\be
	\dualJ_{\star,\kappa}[0,j] = 
	-\ln
	\biggl\{
		\exp
		\biggl(
			-\hf 
			\fder{}{\bar{j}}
			\cdot
				\frac{ \ep_{\kappa}}{1-\rho \ep_{\kappa}}
			\cdot
			\fder{}{\bar{j}}
		\biggr)
		e^{-\homogpr[\bar{j}]}
	\biggr\}.
\label{eq:dualJ-firstguess}
\ee
It is worthwhile recasting this expression. First, let us introduce $\overline{\homog}$ which
has a similar expansion to $\homog$, but with
\be
	\overline{\homog}_{n}(p_1,\ldots,p_n) = \frac{\homogv{n}(p_1,\ldots,p_n)}{p_1^2 \cdots p_n^2},
\qquad
\Rightarrow
\qquad
\homog[\bar{j}] = \overline{\homog}[j],
\quad \homogpr[\bar{j}] = \overline{\homogpr}[j].
\label{eq:overline}
\ee
Now (making explicit certain momentum arguments) we can write
\be
	\dualJ_{\star,\kappa}[0,j] = 
	-\ln
	\biggl\{
		\exp
		\biggl[
			-\hf 
			\int_p
			\fder{}{j(p)}
				\frac{ p^2 \cutoff(p^2/\kappa^2)}{1-\overline{\rho}(p^2) 
				\cutoff(p^2/\kappa^2)}
			\fder{}{j(-p)}
		\biggr]
		e^{-\overline{\homogpr}[j]}
	\biggr\}.
\label{eq:dualJ-firstguess-recast}
\ee

Let us now make the following observation:  if we define new variables $\check{p} \equiv p/\kappa$, $\check{j}(\check{p}) = {j} (p) \kappa^{(d-2+\eta_\star)/2}$, then
\be
	\left.\flowk\right\vert_{\check{j}} \overline{\homogpr}[\check{j}] = 0,
\label{eq:good-flow}
\ee 
Similarly to before, we understand that the partial derivative in~\eq{good-flow} can be taken under the integrals over $\check{p}_i$. Now, if we perform this change of variables in~\eq{dualJ-firstguess}, then we are reasonably close to our aim of finding variables for which the \rhs\ vanishes when differentiated \wrt\ $k$ with said variables held constant.
 However,  there is a problem associated with the operator which hits $e^{-\overline{\homogpr}}$: our change of variables does not make this independent of $k$. Although the (explicit) $\kappa$-dependence of $\cutoff(p^2/\kappa^2) = \cutoff(\check{p}^2)$ disappears, it is reintroduced via $\rho(p^2)$ and the anomalous scaling of $j$. To cure this ill, we must modify~\eq{dual_k}.

\subsubsection{The Second Attempt}
\label{sec:MPE-Second}

The refinement of our method starts by tweaking the first equation of~\eq{dual_k}:
\be
	-\dual'_{\star,\kappa}[\dfield] =
	\ln
	\Bigl(
		e^{\op_{\kappa}} 
		e^{-\dual_\star[\dfield] +\hf \dfield \cdot g \cdot \dfield}
	\Bigr),
\label{eq:dual'}
\ee
where $g = g(p^2;\kappa)$. As we will see below, $g$ will be chosen such that it diverges as $\kappa \rightarrow 0$, meaning that $\lim_{\kappa \rightarrow 0} \dual'_{\star,\kappa} \neq \dual_\star$. However, it will become apparent that $k$ nevertheless plays the role of an IR regulator,  whose effects vanish as $\kappa \rightarrow 0$, when we consider the correlation functions. Putting this issue to one side for the moment, \eq{dual'} implies that the analogue of~\eq{dual_star_k-expr} is
\be
	e^{-\dual'_{\star,\kappa}[\dfield]} = 
	 e^{-\hf \dfield \cdot (\rho-g) / [1-(\rho-g) \, \ep_{\kappa}] \cdot \dfield }
		\exp
		\biggl[
			-\hf 
			\fder{}{\tilde{\dfield}_g} 
			\cdot
				\frac{ \ep_{\kappa}}{1 - (\rho-g) \, \ep_{\kappa}}
			\cdot
			\fder{}{\tilde{\dfield}_g}
		\biggr]
	e^{-\homogpr[\tilde{\dfield}_g]},
\label{eq:analogue}
\ee
where 
\be
	\tilde{\dfield}_g = \dfield/ [1-(\rho-g) \ep_{\kappa}].
\label{eq:dfield_g}
\ee 
Next, let us suppose that
\be
	-\dualJ'_{\star,\kappa}[\dfield,j]
	=
	\ln
	\Bigl(
		e^{-\hf \bar{j} \cdot \omega \cdot \bar{j}}
		e^{\op_{\kappa}} 
		e^{-\dualJ_\star[\dfield,j] + e^{\bar{j} \cdot \delta/\delta \dfield} \hf \dfield \cdot g \cdot \dfield}
	\Bigr),
\label{eq:dual'_star_k-expr}
\ee
with $\omega = \omega(p^2;\kappa)$ to be chosen in a moment. Substituting for $\dualJ_\star[\dfield,j]$ using~\eq{FP-Relationship} we find, employing~\eq{dual'}, that
\be
	\dualJ'_{\star,\kappa}[\dfield,j] = 
	e^{\jbar \cdot (1-\rho \, \ep_{\kappa} ) \cdot \delta / \delta \dfield} 
	\, \dual'_{\star,\kappa}[\dfield]
	- \jbar \cdot \rho \cdot \dfield  
	+\hf \jbar \cdot \bigl[\omega - \rho \bigl(1-\rho \, \ep_{\kappa}\bigr) \bigr]\cdot \jbar,
\label{eq:dualJ-dual-fp-relation'}
\ee
whereupon, substituting in~\eq{analogue} yields
\begin{multline}
	\dualJ'_{\star,\kappa}[\dfield,j] = 
		\hf \bar{j} 
		\cdot
			\biggl[\omega - \rho \bigl(1-\rho \, \ep_{\kappa}\bigr) 
			+ \frac{(\rho-g) (1-\rho \, \ep_{\kappa})^2}{1-(\rho-g) \ep_{\kappa}}
			\biggr]
		\cdot \bar{j}
\\
		-\ln
		\biggl\{
		e^{\jbar \cdot (1-\rho \, \ep_{\kappa} ) \cdot \delta / \delta \dfield} 
		\exp
		\biggl[
			-\hf 
			\fder{}{\tilde{\dfield}_g} 
			\cdot
				\frac{ \ep_{\kappa}}{1 - (\rho-g) \, \ep_{\kappa}}
			\cdot
			\fder{}{\tilde{\dfield}_g}
		\biggr]
		e^{-\homogpr[\tilde{\dfield}_g]}
		\biggr\} +\ldots,
\end{multline}
where the ellipsis represents terms which have at least one power of $\dfield$.
Now, if we choose
\be
	\omega = \frac{g (1-\rho \, \ep_{\kappa} )}{1 - (\rho-g)\ep_{\kappa}}
\ee
then the first term vanishes. Noticing from~\eq{dfield_g} that
\[
	(1-\rho \, \ep_{\kappa} ) \fder{}{\dfield}
	= \frac{1-\rho \, \ep_{\kappa} }{1-(\rho-g) \ep_{\kappa}}
	 \fder{}{\tilde{\dfield}_g},
\]
it is apparent that
\be
	\dualJ'_{\star,\kappa}[0,j] = 
	-\ln
	\biggl\{
		\exp
		\biggl[
			-\hf
			\fder{}{\bar{j}_g}
			\cdot
				\frac{ \ep_{\kappa}}{1-(\rho-g) \ep_{\kappa}}
			\cdot
			\fder{}{\bar{j}_g}
		\biggr]
		e^{-\homogpr[\bar{j}_g]}
	\biggr\},
\label{eq:dualJ'-0}
\ee
where [recalling that $\bar{j}(p) = j(p)/p^2$]
\be
	j_g = \frac{1-\rho\ep_{\kappa}}{1-(\rho-g) \ep_{\kappa}}  {j}.
\label{eq:j'}
\ee
Given a function, $F(p^2/\kappa^2)$---about which we will say more in a moment---the point of all of this can be seen upon choosing%
\be
	g(p^2) = \rho(p^2) -
	 \frac{
	 	1 - \kappa^{\eta_\star-2} p^2 \cutoff(p^2/\kappa^2) F^{-1}(p^2/\kappa^2)
	}{\ep_{\kappa}(p^2)},
\label{eq:g}
\ee
so that, if we identify $W'_{\star,\kappa}[j_g] \equiv -\dualJ'_{\star,\kappa}[0,j] $, it is apparent that we have
\be
	W'_{\star,\kappa}[j_g]  =
	\ln 
	\biggl\{
		\exp
		\biggr[	
			- 
			\kappa^{2-\eta_\star}
			\hf
			\int_p
			\fder{}{j_g(p)}
				F(p^2/\kappa^2)
			\fder{}{j_g(-p)}
		\biggl]	
		e^{-\overline{\homogpr}[j_g]}
	\biggr\}.
\label{eq:W_star_k-desired}
\ee
If we again work with momenta $\check{p} \equiv p/\kappa$ and take $\mathscr{J}(\check{p}) = j_g(p) \kappa^{(d-2+\eta_\star)/2}$ then, using~\eq{fder-dimless} adapted to the case in hand, it is clear that $\delta/\delta \!\mathscr{J}(\check{p}) = \kappa^{(d+2-\eta_\star)/2}\delta / \delta j_g(p) $. Finally, we have achieved our goal: for if we use these variables then, precisely as desired, we have that
\be
	-\left.\flowk\right\vert_{\!\mathscr{J}} \mathscr{W}_{\star,\star}[\mathscr{J}] = 0,
\ee
where $\mathscr{W}_{\star,\star}[\mathscr{J}]  = W'_{\star, \kappa}[j_g]$. Henceforth, we will use the abbreviation $\mathscr{W}_\star \equiv \mathscr{W}_{\star,\star}$. 

Let us now deduce some properties of $F$. First of all, for small $p^2/\kappa^2$, it must exhibit quasi-locality.
Secondly, we require that $\kappa$ plays the role of an IR regulator in~\eq{W_star_k-desired}. Presuming, as before, that the limit $k \rightarrow 0$ can  be taken in the \naive\ way, we can achieve this by demanding that
\be
	F(p^2/\kappa^2) \sim \Bigl(\frac{p^2}{\kappa^2}\Bigr)^{1-\eta_\star/2} \tilde{\cutoff}(p^2/\kappa^2),
	\qquad
	\mathrm{for}\ \kappa \rightarrow 0,
\ee
where $\tilde{\cutoff}$ is some cutoff function which can, in principle, differ from $\cutoff$.%
\footnote{%
It is tempting to suppose that, since $\tilde{\cutoff}(p^2/\kappa^2)$ can be taken to die off faster than any power for large $p^2/\kappa^2$, we are free to ignore the overall $\kappa^{2-\eta_\star}$ in~\eq{W_star_k-desired} when considering the $k \rightarrow 0$ limit. The fallacy of this is readily illustrated by considering $\int^{\infty}_{-\infty} dx \,e^{-x^2/a^2}$: the integrand dies of exponentially fast for small $a$, but the integral dies off only as a power.
}
 This behaviour is consistent with that found in~\cite{HO-Remarks} using a different approach.
Thus, in~\eq{W_star_k-desired}, we see that whatever the value of $\eta_\star$, the limit $k \rightarrow 0$ (with $j_g$ held constant) kills the operator in the big square brackets.
Consequently, $k$ does indeed play the role of an IR regulator, as it must. Indeed, we can now see why it was useful to define $\homogpr$ in~\eq{homogpr}: for if we send $k \rightarrow 0$ in~\eq{W_star_k-desired} then we reproduce the expressions for the correlation functions~\cite{Fundamentals,OJR-Pohl}, including for $\eta_\star =0$.

Before moving on, note that for $\eta_\star =0$ we should take $F(\check{p}^2) = \check{p}^2 \cutoff(\check{p}^2)/[1-\cutoff(\check{p}^2)]$; this satisfies the requirements given above and it is simple to check that things reduce to the fixed-point version of what we did in \sect{Pol}.

Now that we have arranged things such that fixed-points can be readily picked out by a natural criterion applied \wrt\ the IR cutoff, $k$, we can derive a flow equation for the Legendre transform of  $\mathscr{W}$ which inherits the same property. The first thing to do is to rewrite~\eq{W_star_k-desired}
according to
\be
	W'_{\star,\kappa}[j_g] 
	=
	\ln 
	\biggl\{
		\exp
		\biggr(	
			- 
			\hf
			\fder{}{j_g}
			\cdot
				E_{\kappa}
			\cdot
			\fder{}{j_g}
		\biggl)	
		e^{-\overline{\homogpr}[j_g]}
	\biggr\},
\label{eq:W-better}
\ee
where we take
\be
	E_{\kappa}(\check{p}^2) =
	\kappa^{2-\eta_\star}
	F(\check{p}^2).
%
%
\ee
Differentiating~\eq{W-better} \wrt\ $k$ whilst holding $j_g$ constant yields 
an equation almost identical to~\eq{flow-W}:
\be
	\flowk W'_{\star,\kappa}[j_g]
	=
	\hf \classicaljpr{W'_{\star,\kappa}}{\dot{E}_{\kappa}}{W'_{\star,\kappa}}
	+\hf \quantumjpr{\dot{E}_{\kappa}}{W'_{\star,\kappa}},
\label{eq:flow-W-modified}
\ee
where it is apparent that
\be
	\dot{E}_{\kappa}(\check{p}^2) 
	= \kappa^{2-\eta_\star} f(\check{p}^2),
	\qquad
	\mathrm{with}
	\qquad
	 f(\check{p}^2) = 
	\bigl(\eta_\star -2\bigr) F (\check{p}^2)
	+ 2 \check{p}^2 \der{F(\check{p}^2) }{\check{p}^2}.
\ee
Changing variables in~\eq{flow-W-modified} to $\check{p}_i$ and $\mathscr{J}$ we find that
\be
	 -\int_{\check{p}}
	\biggl[
		\mathscr{J}(\check{p})
		\biggl( \frac{\D+ 2 - \eta_\star}{2} + \check{p} \cdot \partial_{\check{p}} \biggr) 
		\fder{}{\!\mathscr{J}(\check{p})}
	\biggr]
	\mathscr{W}_\star[\mathscr{J}]
	=
	\hf \classicalscrJ{\mathscr{W}_\star}{f}{\mathscr{W}_\star}
	+\hf \quantumscrJ{f}{\mathscr{W}_\star}.
\label{eq:flow-W-rescaled}
\ee
Having made clear the essential role played by rendering variables dimensionless using $k$, we will now drop the $\check{}\,$s. Indeed, in~\eq{flow-W-rescaled} $\check{p}$ is anyway a dummy symbol and, in what follows, it should be clear from the context which rescalings have been done.

Now all we need to do is mimic the derivation of the flow equation~\eq{Morris-1PI}. First we define
\be
	\Phi_{\!\!\mathscr{J}}({p}) \equiv \fder{\mathscr{W}_\star[\mathscr{J}]}{\mathscr{J}(-{p})}
\ee
and then adjust $\mathscr{J}({p})$ to $\mathscr{J}_\Phi({p})$ such that $\Phi_{\!\!\mathscr{J}}({p}) = \Phi({p})$. Next we introduce
\be
	\Gtot_\star[\Phi] \equiv \mathscr{J}_\Phi \cdot \Phi -\mathscr{W}_\star[\mathscr{J}_\Phi]
\label{eq:Gtot_star}
\ee
and then make use of
\[
\Phi = \fder{\mathscr{W}_\star[\mathscr{J}_\Phi]}{\!\mathscr{J}_\Phi},
\qquad
\mathscr{J}_\Phi = \fder{\Gtot_\star[\Phi]}{\Phi}
,
\qquad
\int_{{q}} \dfder{\Gtot_\star[\Phi]}{\Phi({p})}{\Phi({q})} 
	\dfder{\mathscr{W}_\star[\mathscr{J}_\Phi]}{\!\mathscr{J}_\Phi({-q})}{\!\mathscr{J}_\field({-p}')}
	= \deltahat{{p}-{p}'},
\]
ultimately obtaining Morris' rescaled fixed-point equation for the effective average action\footnote{The precise identification occurs as follows. Labelling Morris' additive IR cutoff function as $\cutoff_\mathrm{add}$ then,
for a multiplicative IR cutoff function, $\cutoff_{\mathrm{IR}}$, we have $\cutoff^{-1}_\mathrm{add} + 1 = \cutoff^{-1}_{\mathrm{IR}}$. If we identify $\cutoff_{\mathrm{IR}} = 1 - \cutoff$, then this
implies that $\ep_{\mathrm{add}}({p}^2) \equiv \cutoff_\mathrm{add}({p}^2) /{p}^2= F^{-1}({p}^2)$. Noting that $\ep_{\mathrm{add}}$ is equivalent to Morris' $\ep$,
equivalence of~\eq{1PI-flow-rescaled} with Morris' equation is now obvious.}
\be
	 \int_{{p}}
	\biggl[
		\Phi({p})
		\biggl( \frac{\D- 2 + \eta_\star}{2} + {p} \cdot \partial_{{p}} \biggr) 
		\fder{}{\Phi({p})}
	\biggr]
	\Gamma_\star[\Phi]
	=
	\hf \Tr
	\Bigl\{
		f
		\Bigl[
			F+ \Gamma_\star^{(2)}
		\Bigr]^{-1}
	\Bigr\},
%
\label{eq:1PI-flow-rescaled}
\ee
where
\be
	 \Gamma_\star[\Phi] \equiv \Gtot_\star[\Phi] - \hf \Phi \cdot F \cdot \Phi.
\label{eq:decomp}
\ee

\section{Equivalent Fixed-Points}
\label{sec:line}

\subsection{The General Case}

The starting point of the above analysis is a critical fixed-point solution, $\Sint_\star[\dfield]$, of the modified Polchinski equation. However, we know that all such solutions belong to a line of equivalent fixed-points, as in~\eq{line}. We would now like to know how the above analysis changes as we move along this line. To this end, we recall from~\cite{OJR-Pohl} that
\be
	\Sint_\star[\dfield](b_0) \mapsto \Sint_\star[\dfield](b) 
	= e^{a\hat{\Count}} \Sint_\star[\dfield](b_0)
	\qquad
	\Rightarrow
	\qquad
	\dual_\star[\dfield](b_0)  \mapsto \dual_\star[\dfield](b) = 
	e^{a\hat{\Count}} \dual_\star[\dfield](b_0)
\ee
where, as before, $b = b_0+a$. Before moving on, let us pause to note a subtlety. The line of fixed-points generated in this way are only equivalent if either $\eta_\star \neq 0$ or we are at the Gaussian fixed-point. Whilst this seems to imply that non-Gaussian fixed-points with $\eta_\star=0$ are excluded from our analysis this is effectively not the case: so long as we restrict ourselves to theories for which the connected two-point correlation function is positive definite then as shown in~\cite{OJR-Pohl}, the only fixed-point with $\eta_\star=0$ is the Gaussian one.

Returning to our analysis, we note from~\cite{OJR-Pohl} that $e^{a\hat{\Count}} \hf \dfield \cdot h \cdot \dfield = 0$. Recalling~\eqs{homog}{h}, it therefore follows that if we define
\be
	\homogpr[\dfield;a] \equiv \homog[\dfield e^{a/2}] - \frac{c_{\eta_\star}}{2} 
	\int_p \dfield(p) \dfield(-p) p^{2(1+ 
	\eta_\star/2)}
\label{eq:G_a}
\ee
then
\be
	e^{a\hat{\Count}} \dual_\star[\dfield](b_0)
	= \homogpr[\dfield;a] + \hf \dfield \cdot \rho \cdot \dfield.
\ee
In turn, this implies that~\eq{W-better} simply becomes
\be
	W'_{\star,\kappa}[j_g;a] 
	=
	\ln 
	\biggl\{
		\exp
		\biggr(	
			- 
			\hf
			\fder{}{j_g}
			\cdot
				E_{\kappa}
			\cdot
			\fder{}{j_g}
		\biggl)	
		e^{-\overline{\homogpr}[j_g;a]}
	\biggr\}
\label{eq:W-better'}
\ee
and so, after transferring to variables rendered dimensionless using $k$,
we have
\be
	\mathscr{W}_{\star}[\mathscr{J};a] = 
	\ln
	\biggl\{
		\exp
		\biggr(	
			- 
			\hf
			\fder{}{\!\mathscr{J}}
			\cdot
				F
			\cdot
			\fder{}{\!\mathscr{J}}
		\biggl)	
		e^{-\overline{\homogpr}[\mathscr{J};a]}
	\biggr\}.
\label{eq:W_astar}
\ee
Thus we have found that moving along a line of equivalent Wilsonian effective action fixed-points induces us to move along a line of equivalent $\mathscr{W}_{\star}[\mathscr{J};a]$s. 

Now we construct the effective average action. Mimicking our earlier approach, we define
\be
	\Phi_{a \!\mathscr{J}}({p}) \equiv \fder{\mathscr{W}_{\star}[\mathscr{J};a]}{\!\mathscr{J}(-{p})}
\ee
and consider adjusting $\mathscr{J}$ to $\mathscr{J}_{a\Phi}$ such that $\Phi_{a \!\mathscr{J}}$ takes \emph{the same} prescribed form as before \ie\ $\Phi_{a\!\mathscr{J}}({p}) = \Phi({p})$. Next we define the effective average action
according to
\be
	\Gtot_{\star}[\Phi;a] \equiv
	\mathscr{J}_{a \Phi} \cdot \Phi -\mathscr{W}_{\star}[\mathscr{J}_{a\Phi};a]
\label{eq:Gamma_a},
\ee
from which it follows that $\Gtot_{\star}[\Phi;a]$ satisfies precisely the same flow equation as $\Gtot_\star[\Phi]$. Taking
\be
	 \Gamma_{\star}[\Phi;a] \equiv \Gtot_{\star}[\Phi;a] - \hf \Phi \cdot F \cdot \Phi
\label{eq:decomp'}
\ee
then, in turn,
$\Gamma_{\star}[\Phi;a]$ satisfies precisely the same flow equation as $\Gamma_\star[\Phi]$:
\be
	 \int_{{p}}
	\biggl[
		\Phi({p})
		\biggl( \frac{\D- 2 + \eta_\star}{2} + {p} \cdot \partial_{{p}} \biggr) 
		\fder{}{\Phi({p})}
	\biggr]
	\Gamma_{\star}[\Phi;a]
	=
	\hf \Tr
	\Bigl\{
		f
		\Bigl[
			F + \Gamma_{\star}^{(2)}(a)
		\Bigr]^{-1}
	\Bigr\},
\label{eq:1PI-flow-rescaled_a}
\ee
Therefore, the line of equivalent Wilsonian effective actions induces a line of equivalent effective average actions. 

The final step is to understand how $\Gamma_{\star}[\Phi;a]$ depends on $a$. Bearing in mind our earlier comments, we will analyse this question first in the case of $\eta_\star \neq 0$ before treating the Gaussian fixed-point on its own.

\subsubsection{$\eta_\star \neq 0$}

In this section we will derive a closed expression for $\Gamma_{\star}[\Phi;a]$ in terms of $\Gamma_\star[\Phi]$. However, before doing so we will use a simple method to derive the $\order{a}$ result. Not only will this serve as a crosscheck for our general result, but also immediately gives us the form of the marginal, redundant operator which generates the line of fixed-points.
To do this, let us define
\be
	\delta  \! \mathscr{J}_{\Phi} ({p}) \equiv \mathscr{J}_{a\Phi}({p}) - \mathscr{J}_{\Phi}({p}),
\ee
where, for small $a$,  $ \delta  \!\mathscr{J}_{\Phi}({p}) = \order{a}$. We can thus rewrite~\eq{Gamma_a} according to
\be
	\Gtot_{\star}[\Phi;a] = \Gtot_{\star}[\Phi] 
	+\mathscr{W}_{\star}[\mathscr{J}_{\Phi}]
	-\mathscr{W}_{\star}[\mathscr{J}_{a\Phi};a]
	+ \delta  \! \mathscr{J}_{\Phi} \cdot \Phi,
\ee
from which it follows that
\begin{multline}
	\Gamma_{\star}[\Phi;a] = \Gamma_{\star}[\Phi] 
	+\mathscr{W}_{\star}[\mathscr{J}_{\Phi}]
	-\mathscr{W}_{\star}[\mathscr{J}_{\Phi};a]
\\
	+ \delta  \! \mathscr{J}_{\Phi} \cdot \fder{}{\mathscr{J}_\Phi}
	\bigl(
		\mathscr{W}_\star[\mathscr{J}_\Phi] - \mathscr{W}_{\star}[\mathscr{J}_{\Phi};a]
	\bigr)
	-
	\hf
	\int_{p,q}
	\delta\! \mathscr{J}_\Phi(p)\, \delta\!\mathscr{J}_\Phi(q)
	\dfder{\mathscr{W}_{\star}[\mathscr{J}_\Phi;a]}{\!\mathscr{J}_\Phi(p)}{\!\mathscr{J}_\Phi(q)}
	-\cdots
\end{multline}
At $\order{a}$, only the first line contributes; therefore, to this order, we require only an expression for $\mathscr{W}_{\star}[\mathscr{J}_{\Phi};a]$ and not an expression for $\delta\! \mathscr{J}_\Phi$.

To proceed, let us focus on~\eq{W_astar}. For $\eta_\star \neq 0$ it follows from~\eqs{h}{G_a} that $\overline{\homogpr}[\mathscr{J};a] = \overline{\homog}[\mathscr{J}e^{a/2}]$. Consequently, each external $\mathscr{J}$ comes with a factor of $e^{a/2}$, whereas each internal line comes with a factor of $e^{a/2}$ at each end. Therefore, we can write
\begin{subequations}
\begin{align}
	\mathscr{W}_{\star}[\mathscr{J};a] 
	&= 
	\ln
	\biggl\{
		\exp \biggl( \frac{a}{2} \mathscr{J} \cdot \fder{}{\!\mathscr{J}} \biggr)	
		\exp
		\biggr(	
			- 
			\frac{e^a}{2}
			\fder{}{\!\mathscr{J}}
			\cdot
				F
			\cdot
			\fder{}{\!\mathscr{J}}
		\biggl)	
		e^{-\overline{\homogpr}[\mathscr{J}]}
	\biggr\}
\label{eq:W_astar-G}
\\
	&
	=
	\ln
	\biggl\{
		\exp \biggl( \frac{a}{2} \mathscr{J} \cdot \fder{}{\!\mathscr{J}} \biggr)
		\exp
		\biggr(	
			\frac{1-e^a}{2}
			\fder{}{\!\mathscr{J}}
			\cdot
				F
			\cdot
			\fder{}{\!\mathscr{J}}
		\biggl)	
		e^{\mathscr{W}_\star[\mathscr{J}]}
	\biggr\}.
\label{eq:W_a*-W_*}
\end{align}
\end{subequations}
Setting $\mathscr{J} = \mathscr{J}_\Phi$ and expanding to $\order{a}$, the result is particularly simple:
\be
	\mathscr{W}_{\star}[\mathscr{J};a] 
	=\mathscr{W}_{\star}[\mathscr{J}] 
	+\frac{a}{2}
	\biggl\{
		\Phi \cdot \fder{\Gamma_\star}{\Phi}
		-
		\Tr
		\Bigl(
			F\bigl[F+ \Gamma_\star^{(2)}\bigr]^{-1}
		\Bigr)
	\biggr\} + \order{a^2}.
\ee
From this it follows that
\be
	\Gamma_{\star}[\Phi;a] 
	= \Gamma_{\star}[\Phi]
	-\frac{a}{2}
	\biggl\{
		\Phi \cdot \fder{\Gamma_\star}{\Phi}
		-
		\Tr
		\Bigl(
			\bigl[1+ F^{-1} \Gamma_\star^{(2)}\bigr]^{-1}
		\Bigr)
	\biggr\}
	+\order{a^2},
\label{eq:line-O(a)}
\ee
allowing us to directly read off the expression for the marginal, redundant operator which generates the line of equivalent fixed-points, in agreement with~\cite{HO-Remarks}.

Having obtained this result, we now turn to the general treatment. Our starting point is the standard result~\cite{WeinbergII} 
\be
	\mathscr{W}_\star[X]
	= \raisebox{-3ex}{$\stackrel{\ds \int}{\mathrm{\scriptstyle connected\ tree}}$}
	\measure{\Phi}
	e^{-\Gtot_\star[\Phi] - X \cdot \Phi},
\label{eq:standard}
\ee
where the functional integral is performed with $X$ held constant.
We now rewrite this according to
\begin{align}
	\mathscr{W}_\star[X]
	& = \raisebox{-3ex}{$\stackrel{\ds \int}{\mathrm{\scriptstyle connected\ tree}}$}
	\measure{\Phi}
	e^{-\hf \Phi \cdot F \cdot \Phi -\Gamma_\star[\Phi] - X \cdot \Phi}
\nonumber
\\
	& = 
	\ln
	\biggl\{
		\exp
		\biggl(
			\hf \fder{}{\Phi} \cdot F^{-1} \cdot \fder{}{\Phi} 
		\biggr)_{\!\mathrm{t}}
		e^{-\Gamma_\star[\Phi] - X \cdot \Phi}
	\biggr\}_{\Phi=0}
\nonumber
\\
	& = 
	\hf X \cdot F^{-1} \cdot X +
	\ln
	\biggl\{
		\exp
		\biggl(
			\hf \fder{}{X} \cdot F \cdot \fder{}{X} 
		\biggr)_{\!\mathrm{t}}
		e^{-\Gamma_\star[F^{-1} X]}
	\biggr\},
\end{align}
where the subscript `t' instructs us to keep only the tree graphs generated by the associated operator.
It follows that
\be
	\mathscr{W}_{\star}[\mathscr{J}_\Phi;a]
	=
	\hf \mathscr{J}_\Phi \cdot F^{-1} \cdot \mathscr{J}_\Phi +
	\ln
	\biggl\{
		\exp
		\biggl(
			\hf \fder{}{\!\mathscr{J}_\Phi} \cdot F \cdot \fder{}{\!\mathscr{J}_\Phi} 
		\biggr)_{\!\mathrm{t}}
		e^{-\Gamma_{\star}[F^{-1}\mathscr{J}_\Phi;a]}
	\biggr\}.
\label{eq:W_astar-TreeExpansion}
\ee
One of the nice things about this representation of $\mathscr{W}_{\star}[\mathscr{J}_\Phi;a]$ is that we can invert to find $\Gamma_{\star}[F^{-1}\mathscr{J}_\Phi;a]$, as follows from~\cite{NonRenorm}: 
\be
	\Gamma_{\star}[F^{-1}\mathscr{J}_\Phi;a] =
	-
	\ln
	\biggl\{
		\exp
		\biggl(
			-\hf \fder{}{\!\mathscr{J}_\Phi} \cdot F \cdot \fder{}{\!\mathscr{J}_\Phi} 
		\biggr)_{\!\mathrm{t}}
		e^{\mathscr{W}_{\star}[\mathscr{J}_\Phi;a]-\hf \mathscr{J}_\Phi \cdot F^{-1} \cdot \mathscr{J}_\Phi}
	\biggr\}.
\label{eq:invert}
\ee
Utilizing~\eq{W_a*-W_*} with $\mathscr{J} = \mathscr{J}_\Phi$ we obtain
\begin{multline}
	\Gamma_{\star}[F^{-1}\mathscr{J}_\Phi;a] 
	=
	-\ln
	\biggl\{
		\exp
		\biggl(
			-\hf \fder{}{\!\mathscr{J}_\Phi} \cdot F \cdot \fder{}{\!\mathscr{J}_\Phi} 
		\biggr)_{\!\mathrm{t}}
\\
	e^{-\hf \mathscr{J}_\Phi \cdot F^{-1} \cdot \mathscr{J}_\Phi}
	\exp \biggl( \frac{a}{2} \mathscr{J}_\Phi \cdot \fder{}{\!\mathscr{J}_\Phi} \biggr)	
	\exp
		\biggr(	
			\frac{1-e^a}{2}
			\fder{}{\!\mathscr{J}_\Phi}
			\cdot
				F
			\cdot
			\fder{}{\!\mathscr{J}_\Phi}
		\biggl)
	e^{\mathscr{J}_\Phi \cdot \Phi -\Gtot_\star[\Phi]}
	\biggr\}.
\end{multline}
(Note that the action of the tree-level operator on objects which already contain loop integrals is simply defined such that it does not change the number of loops.)
Next, define a new field $\mathscr{Y}_\Phi({p}) \equiv F^{-1}({p}^2) \mathscr{J}_\Phi({p})$, so that we have
\begin{multline}
	\Gamma_{a\star}[\mathscr{Y}_\Phi] 
	=
	-\ln
	\biggl\{
		\exp
		\biggl(
			-\hf \fder{}{\mathscr{Y}_\Phi} \cdot F^{-1} \cdot \fder{}{\mathscr{Y}_\Phi} 
		\biggr)_{\!\mathrm{t}}
\\
	e^{-\hf \mathscr{Y}_\Phi \cdot F \cdot \mathscr{Y}_\Phi}
	\exp \biggl( \frac{a}{2} \mathscr{Y}_\Phi \cdot \fder{}{\mathscr{Y}_\Phi} \biggr)	
	\exp
		\biggr(	
			\frac{1-e^a}{2}
			\fder{}{\mathscr{Y}_\Phi}
			\cdot
				F^{-1}
			\cdot
			\fder{}{\mathscr{Y}_\Phi}
		\biggl)
	e^{\mathscr{Y}_\Phi \cdot F \cdot \Phi -\Gtot_\star[\Phi]}
	\biggr\}.
\label{eq:Gamma_star[Y]}
\end{multline}
Setting $a=0$ produces
\be
	\Gamma_{\star}[\mathscr{Y}_\Phi] 
	=
	-\ln
	\biggl\{
		\exp
		\biggl(
			-\hf \fder{}{\mathscr{Y}_\Phi} \cdot F^{-1} \cdot \fder{}{\mathscr{Y}_\Phi} 
		\biggr)_{\!\mathrm{t}}
		e^{-\hf \mathscr{Y}_\Phi \cdot F \cdot \mathscr{Y}_\Phi+\mathscr{Y}_\Phi \cdot F \cdot \Phi -\Gtot_\star[\Phi]}
	\biggl\};
\ee
inverting and substituting for $\mathscr{Y}_\Phi \cdot F \cdot \Phi -\Gtot_\star[\Phi]$ in~\eq{Gamma_star[Y]} yields an expression for the line of equivalent fixed-points entirely in terms of $\Gamma_\star$; since all functionals now depend on $\mathscr{Y}_\Phi$, we will change this (dummy) symbol to $\Phi$:
\begin{multline}
	\Gamma_{\star}[\Phi;a] 
	=
	-\ln
	\biggl\{
		\exp
		\biggl(
			-\hf \fder{}{\Phi} \cdot F^{-1} \cdot \fder{}{\Phi} 
		\biggr)_{\!\mathrm{t}}
	e^{-\hf \Phi \cdot F \cdot \Phi}
	\exp \biggl( \frac{a}{2} \Phi \cdot \fder{}{\Phi} \biggr)	
\\
	\exp
		\biggr(	
			\frac{1-e^a}{2}
			\fder{}{\Phi}
			\cdot
				F^{-1}
			\cdot
			\fder{}{\Phi}
		\biggl)
	e^{\hf \Phi \cdot F \cdot \Phi}	
	\exp
		\biggl(
			\hf \fder{}{\Phi} \cdot F^{-1} \cdot \fder{}{\Phi} 
		\biggr)_{\!\mathrm{t}}
	e^{-\Gamma_\star[\Phi]}
	\biggr\}.
\label{eq:final}
\end{multline}
Thus, given a fixed-point solution $\Gamma_\star$, this equation can be used to generate the line of
equivalent fixed-points, $\Gamma_{a\star}$.

We can check consistency with our previous result~\eq{line-O(a)} by expanding to $\order{a}$. Using the result that (up to a discarded vacuum energy term),
\be
	\biggl[
		\Phi \cdot \fder{}{\Phi} - \fder{}{\Phi} \cdot F^{-1} \cdot \fder{}{\Phi} 
		,
		e^{\hf \Phi \cdot F \cdot \Phi}
	\biggr]
	=
	-2 e^{\hf \Phi \cdot F \cdot \Phi} \Phi \cdot \fder{}{\Phi},
\ee
it is straightforward to show that
\begin{multline}
	\Gamma_{\star}[\Phi;a] + \order{a^2} = \Gamma_{\star}[\Phi] 
\\
	+\frac{a}{2}
	e^{\Gamma_\star[\Phi]}
	\exp
		\biggl(
			-\hf \fder{}{\Phi} \cdot F^{-1} \cdot \fder{}{\Phi} 
		\biggr)_{\!\mathrm{t}}
	\biggl[
		\Phi \cdot \fder{}{\Phi} +
		\fder{}{\Phi} \cdot F^{-1} \cdot \fder{}{\Phi} 
	\biggr]
	\exp
		\biggl(
			\hf \fder{}{\Phi} \cdot F^{-1} \cdot \fder{}{\Phi} 
		\biggr)_{\!\mathrm{t}}
	e^{-\Gamma_\star[\Phi]}
\end{multline}
Commuting the $\Phi \cdot \delta / \delta \Phi$ through the operator to its right yields
\begin{multline}
	\Gamma_{\star}[\Phi;a] + \order{a^2} = \Gamma_{\star}[\Phi] 
	- \frac{a}{2}\biggl\{ \Phi \cdot \fder{\Gamma_\star}{\Phi}
	-
	e^{\Gamma_\star[\Phi]}
	\exp
		\biggl(
			-\hf \fder{}{\Phi} \cdot F^{-1} \cdot \fder{}{\Phi} 
		\biggr)_{\!\mathrm{t}}
\\
	\times
	\biggl[
		\fder{}{\Phi} \cdot F^{-1} \cdot \fder{}{\Phi} 
		-
		\biggl(
			\fder{}{\Phi} \cdot F^{-1} \cdot \fder{}{\Phi} 
		\biggr)_{\!\mathrm{t}}
	\biggr]
	\exp
		\biggl(
			\hf \fder{}{\Phi} \cdot F^{-1} \cdot \fder{}{\Phi} 
		\biggr)_{\!\mathrm{t}}
	e^{-\Gamma_\star[\Phi]}
	\biggr\}.
\end{multline}
The difference of the operators in the big square brackets yields a single operator which is compelled
to generate a single loop; this will be denoted by the tag `l'. Consider now
\[
	\biggl(
		\fder{}{\Phi} \cdot F^{-1} \cdot \fder{}{\Phi} 
	\biggr)_{\!\mathrm{l}}
	\exp
	\biggl(
		\hf \fder{}{\Phi} \cdot F^{-1} \cdot \fder{}{\Phi} 
	\biggr)_{\!\mathrm{t}}
	e^{-\Gamma_\star[\Phi]}.
\]
The rightmost operator generates all tree diagrams; the leftmost piece ties up part of each tree into a loop.
The sum of all such terms can be simplified by noticing that the entire series can be generated from just
the 1PI diagrams:
\[
	\biggl(
		\fder{}{\Phi} \cdot F^{-1} \cdot \fder{}{\Phi} 
	\biggr)_{\!\mathrm{l}}
	\exp
	\biggl(
		\hf \fder{}{\Phi} \cdot F^{-1} \cdot \fder{}{\Phi} 
	\biggr)_{\!\mathrm{t}}
	e^{-\Gamma_\star[\Phi]}
	=
	\biggl(
		\hf \fder{}{\Phi} \cdot F^{-1} \cdot \fder{}{\Phi} 
	\biggr)_{\!\mathrm{t}}
	e^{-\Gamma_\star[\Phi]}
	\bigr(\mbox{1PI$_1$ diagrams}\bigr),
\]
where the 1PI$_1$ diagrams are one-loop diagrams built from vertices of $-\Gamma_\star$ joined together by instances of $F^{-1}$. But this simply corresponds to the vertex expansion of 
$\Tr
\{
	[1+ F^{-1} \Gamma_\star^{(2)}]^{-1}
\}$,
and so we recover~\eq{line-O(a)}.

\subsubsection{The Gaussian Fixed-Point}

As one might expect, the Gaussian fixed-point is very easy to treat; indeed, we can
derive some results that, in other cases, would be very hard to obtain. So, rather than immediately solving the fixed-point equation~\eq{1PI-flow-rescaled} for a representative of the Gaussian fixed-point and
then mimicking the analysis of the previous section to generate the associated line, we will take a more circumspect approach.
In particular, instead of starting with the effective average action, we will start our analysis with the Wilsonian effective action. 
Using the conventions of previous works~\cite{Fundamentals,OJR-Pohl}, the line of Gaussian fixed-points (for which $\eta_\star =0$) is
\be
	S_\star[\dfield](b) = -\hf \int_p  \dfield(p) \dfield(-p) \frac{e^{b} p^2}{1+ e^{b} \cutoff (p^2)}.
\label{eq:line-Gaussian}
\ee
Taking $b = b_0$ to be a reference point, is easy enough to check~\cite{Fundamentals} that $e^{a\hat{\Count}} S_\star[\dfield](b_0)  = S_\star[\dfield](b_0 + a) $, with $b_0 + a =b$. The result~\eq{line-Gaussian}
corresponds to
\be
	\dual_\star[\dfield](b) = \homog[\dfield](b) = -\frac{e^{b}}{2} \int_p \dfield(p) \dfield(-p) p^2,
\ee
and from this perspective it is clear that $e^b$ plays the role of a normalization constant. 
Recalling~\eq{rho}, \eqs{homog}{h}, notice that the first equality follows because, for $\eta_\star=0$, $h(p^2) = 0$. 
Recalling from~\eq{overline} that $\overline{\homogpr}_a[\dfield] = \homogpr_a[\overline{\dfield}]$, with $\overline{\dfield}(p) = \dfield(p)/p^2$ it is apparent from~\eq{G_a} that, in the current scenario,
\be
	\overline{\homogpr}_a[\dfield] = -\frac{1+e^{b_0+a}}{2} 
	\int_{{p}} \dfield({p}) \dfield(-{p}) \frac{1}{{p}^2},
\ee
where it is now convenient to split up $b = b_0+a$
and so, from~\eq{W_astar}, we have that
\be
	\mathscr{W}_{\star}[\mathscr{J};a] = 
	\hf  \int_{{p}} \mathscr{J}({p}) \mathscr{J}(-{p})
	\frac{1+ e^{b_0+a}}{{p}^2+(1+e^{b_0+a}) F({p}^2)}.
\label{eq:W_a-G}
\ee
From the definition of $\mathscr{J}_{a\Phi}$:
\be
	\fder{\mathscr{W}_{\star}[\mathscr{J};a]}{\mathscr{J}}\biggr\vert_{\mathscr{J} = \mathscr{J}_{a\Phi}} = \Phi
\ee
we immediately see that
\be
	\mathscr{J}_{a\Phi}({p}) = \Phi({p}) \frac{{p}^2 
	+(1+e^{b_0+a}) F({p}^2)}{1+e^{b_0+a}};
\label{eq:J_a-G}
\ee
the Gaussian case is so simple that we have been able to easily find the form of $\mathscr{J}_a$ which
induces $\Phi_{a\mathscr{J}}$ to obtain the reference form $\Phi$.
Substituting~\eqs{W_a-G}{J_a-G} into~\eq{Gamma_a} yields
\be
	\Gtot_{\star}[\Phi;a] = \hf
	\int_{{p}} \Phi({p}) \Phi(-{p})  
	\frac{{p}^2+(1+e^{b_0+a}) F({p}^2)}{1+e^{b_0+a}}
\ee
and so, from~\eq{decomp}, we obtain the result
\be
	\Gamma_{\star}[\Phi;a] 
	= \frac{1}{2 (1+e^{b_0+a})}\int_{{p}} \Phi({p}) \Phi(-{p}) {p}^2.
\label{eq:Gamma-Gline}
\ee
It is trivial to check that this is, indeed, a solution to the fixed-point equation~\eq{1PI-flow-rescaled}
with $\eta_\star =0$.

\section{Conclusion}
\label{sec:conc}

The analysis of this paper has been somewhat involved, and so we now recapitulate the main steps.
To begin with, we started with the plain Polchinski equation, from which it has been known for a long time how to derive (in several different ways) a flow equation for the effective average action, $\Gamma$. Inspired by the approach of Ellwanger~\cite{Ellwanger-1PI}, the standard flow equation for $\Gamma$ was obtained in~\eq{Morris-1PI}, with the minimum of fuss.

However, the plain Polchinski equation is not the most convenient flow equation of the Wilson-Wegner-Polchinski type for discovering fixed-points. This is because the redundant coupling, Z, (the field strength renormalization) explicitly appears in the action. Since this coupling can be removed by a quasi-local field redefinition, there is no need for it to stop flowing at what, for the remaining couplings, is a fixed-point. Therefore, the apparently natural fixed-point criterion $\flow S_\star = 0$ (applied after scaling out the various canonical dimensions) will only pick out solutions for which the anomalous dimension of the field vanishes (the only physically admissible solution of this type is the Gaussian one~\cite{OJR-Pohl}); discovering other fixed-points in this formalism is possible but awkward.

The most natural solution to this problem is to modify the flow equation, by incorporating a particular field redefinition, so that $Z$ is removed from the action. Having done this, the criterion $\flow S_\star = 0$ now has the capacity to find fixed-points with non-zero anomalous dimension.\footnote{The reason why it is likely that further modifying the flow equation to remove other redundant couplings will \emph{not} reveal new fixed-points is discussed in~\cite{Fundamentals}.} However, modifying the flow equation means that the path from $S$
to a flow equation for $\Gamma$ must be rethought.

As in the plain Polchinski equation, the first step is to derive a flow equation for the IR regulated generator of connected correlation functions, $W_k$. However, there is some freedom as to precisely how we define the latter.%
\footnote{This freedom is there even at the level of the Polchinski equation. For example, we could introduce an IR regularization in a different way from~\eq{op-reg}. A simple example would be to replace~\eq{Cutoff-UV-IR} by the difference of two \emph{different} cutoff functions, but with both normalized such that for zero argument they yield unity. The object derived from the Wilsonian effective action along the lines of~\eq{Sint_k} would still correspond to an IR regularized generator of the correlation functions but it would not satisfy~\eq{flow-W}. As such, it would not be very nice to deal with but nevertheless illustrates the freedom in constructing IR regularized generating functionals from the Wilsonian effective action.}
In fact, rather than dealing with the full scale-dependent case, in this paper we focused just on fixed-points. Our aim, then, was to define an appropriate object, $W_{\star, \kappa}$, understood as an IR regularized version of $W_\star$. Our first attempt to do this began with~\eq{dual_k}. Unfortunately, by the time we arrived at~\eq{dualJ-firstguess-recast}, it was apparent that there was a short-coming. 

The seemingly natural thing to have done at this point would be to identify $W_{\star, \kappa}[j]$
with $-\dualJ_{\star, \kappa}[0,j]$. But we placed an additional requirement on our construction, which this identification fails to fulfil. The requirement is as follows. By construction, $W_{\star, \kappa}[j]$ is derived from a fixed-point object, where fixed-point objects are defined such that their derivatives \wrt\ $\Lambda$ vanish. Now, our aim was to pass to a formalism in which no mention of $\Lambda$ is made, and all scale derivatives are \wrt\ the IR scale, $k$. Thus \emph{purely for convenience}, we would like a simple criterion \wrt\ $k$ which tells us, without reference to the construction via a fixed-point Wilsonian effective action, that we are dealing with a fixed-point quantity. The natural criterion is obviously that the scale derivative \wrt\ $k$ vanishes. Thus, in~\eqs{dual'}{dual'_star_k-expr} we refined our guess~\eq{dual_k}; this allowed us to construct a $W_{\star, \kappa}[j]$  which has two important properties:
\begin{enumerate}
	\item It has an interpretation as an IR regularized version of $W_{\star}[j]$;
	
	\item After passing to appropriate variables, its $k$-derivative vanishes.
\end{enumerate}

That we have had to tweak our construction in order to ensure the second property is of no concern. After all, when dealing with the Wilsonian effective action, we tweaked the Polchinski equation in order to be able to use a simple criterion to find fixed-points; and in the case of $W_{\star, \kappa}$ we have followed the same philosophy: our approach is motivated by convenience and not necessity. Having found the desired form for $W_{\star, \kappa}[j]$, we then performed the usual Legendre transform to derive a fixed-point equation, \eq{1PI-flow-rescaled}, for $\Gamma$, recovering Morris' fixed-point equation of~\cite{TRM-Deriv}. Let us note that this is the first time that this equation has been derived from the underlying Wilsonian formalism.

An advantage of finding this link between the two formalisms is that results from one can now be readily mapped to the other. In \sect{line} we exploited this to find expressions for the line of equivalent fixed-points associated with every critical fixed-point; the result for $\eta_\star \neq 0$ is given by~\eq{final}, whereas the result for the Gaussian fixed-point is given by~\eq{Gamma-Gline}. Compared to the corresponding expression for the Wilsonian effective action, \eq{line}, the formula~\eq{final} is rather complicated. Indeed, this seems to further reinforce a general feeling that structural results are most easily obtained in the Wilson-Wegner-Polchinski approach. The flip side of this is that the effective average action seems superior for numerical studies.

In terms of future work, the results of this paper should be straightforward to generalize to the supersymmetric case using the methodology of~\cite{SUSY-Chiral} and to noncommutative theories by appropriately adapting~\cite{RG+OJR}. This should be of relevance in the context of~\cite{Synatschke:2009nm,Gies:2009az} and~\cite{Koslowski-FRG-NCSFT}, respectively. Moreover, it should be reasonably easy to extend the analysis of this paper away from fixed-points, which would provide a direct derivation of Morris' full flow equation of~\cite{TRM-Deriv} from the underlying Wilsonian formalism.

\begin{acknowledgments}
	This work was supported by the Science and Technology Facilities Council 
	[grant number ST/F008848/1]. I would like to thank Hugh Osborn both for
	encouraging me to do this piece of work and also for some very
	useful comments on the manuscript.
\end{acknowledgments}

\bibliography{../../../Biblios/ERG.bib,../../../Biblios/Books.bib,../../../Biblios/NC.bib}

\end{document}